\documentclass{edm_article}

\usepackage{hyperref}           
\usepackage{booktabs}           
\usepackage{amsfonts}           
\usepackage{amsmath}            
\usepackage{amssymb}            
\usepackage{cleveref}           
\usepackage[numbers]{natbib}    
\usepackage{soul}               
\usepackage{graphicx}
\usepackage{multirow}
\usepackage{float}
\usepackage[caption=false]{subfig}
\usepackage{array}
\usepackage{balance}

\newcolumntype{M}[1]{>{\centering\arraybackslash}m{#1}}

\newcommand{\x}{\mathbf{x}}

\newcommand{\ie}{{\em i.e.~\/}}
\newcommand{\eg}{{\em e.g.~\/}}
\newcommand{\etal}{{\em et.~al.~\/}}

\newcommand{\vs}{{\em vs.~\/}}
\newcommand{\cf}{{\em c.f.~\/}}

\renewcommand{\lim}{\operatornamewithlimits{lim}}

\begin{document}

\title{Equitable Ability Estimation in Neurodivergent Student Populations with Zero-Inflated Learner Models}

\Crefname{equation}{Eqn}{Eqns}
\crefname{equation}{Eqn}{Eqns}
\Crefname{figure}{Fig}{Figs}
\crefname{figure}{Fig}{Figs}
\crefname{tabular}{Table}{Tables}
\Crefname{tabular}{Table}{Tables}
\crefname{section}{Sec}{Secs}
\Crefname{section}{Sec}{Secs}



\numberofauthors{5}
\author{
\alignauthor 
            Niall Twomey\titlenote{Corresponding author.}\\
            \affaddr{Kidsloop Ltd, UK}\\
            \email{niall.twomey@kidsloop.live}
\and
\alignauthor 
            Sarah McMullan\\
            \affaddr{auticon, UK; Kidsloop Ltd, UK}\\
            \email{sarah.mcmullan@auticon.co.uk}
\and
\alignauthor 
            Anat Elhalal\\
            \affaddr{Kidsloop Ltd, UK}\\
            \email{anat.elhalal@kidsloop.live}
\and 
\alignauthor Rafael Poyiadzi\\
            \affaddr{Kidsloop Ltd, UK}\\
            \email{rafael.poyiadzi@kidsloop.live}
\and
\alignauthor Luis Vaquero\\
            \affaddr{Kidsloop Ltd, UK}\\
            \email{luis.vaquero@kidsloop.live}
}

\maketitle

\begin{abstract}

At present, the educational data mining community lacks many tools needed for ensuring equitable ability estimation for Neurodivergent (ND) learners. On one hand, most learner models are susceptible to under-estimating ND ability since confounding contexts cannot be held accountable (\eg consider dyslexia and text-heavy assessments), and on the other, few (if any) existing datasets are suited for appraising model and data bias in ND contexts. In this paper we attempt to model the relationships between context (delivery and response types) and performance of ND students with zero-inflated learner models. This approach facilitates simulation of several expected ND behavioural traits, provides equitable ability estimates across all student groups from generated datasets, increases interpretability confidence, and can significantly increase the quality of learning opportunities for ND students. Our approach consistently out-performs baselines in our experiments and can also be applied to many other learner modelling frameworks. 

\end{abstract}


\section{Introduction}
\label{sec:introduction}

In the UK, it is estimated that 15\% of the population are ND, having neurological functions that differ from what is considered typical \citep{Lollini2018}. Neurodiversity covers the range of differences in individual brain function and behavioural traits, regarded as part of normal variation in the human population \citep{Singer1999}. Each Neurodivergent Condition (NDC) uniquely affects how information is absorbed, processed, and communicated \citep{patrick2020memory, boyd2018leveling}. Our objective is to adapt Learner Models (LMs) for the individual requirements of a number of NDCs in learning environments, focusing specifically on dyslexia, dyscalculia and Sensory Processing Disorder (SPD) (with prevalences of 10\%, 6\% and 5-15\% respectively \cite{crisfield1995dyslexia,shalev2000developmental,galiana2020sensory}). 

Achievement gaps due to NDCs occur early in life and persist through adolescence into adulthood \cite{ferrer2015achievement}. In many cases, impeded learning opportunities for ND students result from unsuitable learning contexts or lack of adequate student support rather than intrinsically low student ability \cite{papathoma2020guidance}. However, as learning begins to move further into the digital space \citep{homer2021using, plass2020toward}, LMs, which are statistical models of student attainment, will use historic performance to estimate student ability. Owing to a legacy of potentially poor learning contexts, the ability of ND students tends to be under-estimated by LMs since they are not equipped to distinguish between context- and ability-based explanations of performance. Without deliberate effort, therefore, it is very likely that LMs will become biased and offer inequitable recommendations for ND students. On the other hand, opportunities to quell these achievement gaps before they grow are at hand in smart learning environments if LMs are empowered to reason about alternative explanations of performance. 

LM research is highly active in the Educational Data Mining (EDM) community. State-of-the-art approaches include deep neural networks \cite{piech2015deep,gervet2020deep,pandey2019self}, and nonparametric Bayesian methods \cite{khajah2016deep}. We find that the literature is sparse for inclusive LMs applied to ND populations, and we were unable to find many bespoke models or datasets (real or synthetic) even in recent literature reviews \cite{abyaa2019learner,liu2021survey}. Kohli \etal \cite{kohli2010identifying} introduced an approach for identifying dyslexic students based on historic patterns of behaviour and artificial neural networks. 
Mejia \etal \cite{mejia2017inclusive} approached the task by estimating learner’s cognitive deficit specifically for students with dyslexia or reading difficulties.
Ensuring the equity of LM is an important area of research, 
and learning interfaces can be improved by offering multiple assessment Delivery and Response Type (DRT) \cite{papathoma2020guidance}. 
Other works have elaborated further on scores and metrics for ethical and equitable recommendation systems with broad stakeholders, including dyslexic students \cite{marras2021equality}. 
Equity is also explored along explainability and interpretability axes. Some classical LMs 
are readily interpretable and offer intuitive explanations of datasets \cite{pelanek2017bayesian,mandalapu2021we}, though caution must be exercised to avoid over-interpreting models \cite{holstein2021equity}. 

ND students face at least two additional hurdles than Neurotypical (NT) students in learning environments: 1) their ability is inaccurately modelled due to LMs shortcomings; and 2) choosing the most suitable learning context for them to express their true ability is rarely considered. Furthermore, the EDM community currently lacks datasets and simulation tools for developing LMs and assessing equity for NDC contexts. We address these three limitations in this work, by motivating and defining equitable LMs for ND students (\Cref{sec:methods}), defining a simulation environment (\Cref{sec:methods:sim}), and demonstrating strong performance in our results and conclusions (\Cref{sec:results,sec:conclusions}).

\section{Methods}
\label{sec:methods}

Due to a lack of available datasets that include ND students, we explore equitable estimation in simulations. Our model combines the use of Zero-Inflated Models (ZIMs) \cite{lambert1992zero} and Item Response Theory (IRT) \cite{Barton1981, Liao2012}. 
Our assumption is that DRT choices will affect the quality of learning opportunities for ND students, with unsuitable DRT resulting in lower Learning Quality Factor (LQF). Without considering the suitability of DRTs for students, LMs risk recommending low-quality learning opportunities and mis-interpreting poor performance on these as an indication of low student ability. The model and simulation procedure proposed is designed to be used to identify the best DRTs for each student, and prevent underestimation of abilities. 

\subsection{IRT-based Zero-Inflated Learner Model}
\label{sec:methods:irtzilm}

Our proposed approach, Zero-inflated Learner Models (ZILMs), shown in \Cref{eq:zim}, builds on the assumption that there are two explicit explanations of zeros: 1) low ability relative to difficulty (low $p$); and 2) low LQF (high $\pi$). With this formulation, a zero from a student with high ability with in an unsuitable DRT can be explained by the poor LQF since $\pi$ has high responsibility for the outcome \cite{bishop2006pattern}. 
\begin{align}
    \label{eq:zim}
Pr(Y=y) = &\begin{cases}
    \pi + (1 - \pi) \cdot (1-p) & \text{if } y = 0\\
    (1 - \pi) \cdot p            & \text{if } y = 1
\end{cases}
\end{align}
In our setting, $p$ is based on IRT, and $\pi$ (which reflects LQFs) is parameterised by item, NDC and DRT features (\cf \Cref{sec:methods:sim}), resulting in IRT-based ZILM (IRT-ZILM).

IRT was chosen as the base LM in IRT-ZILM over alternative options as: 1) IRT is well-understood and simple to interpret; 2) Bayesian Knowledge Tracing (BKT) is known to have over- and under-estimation problems \cite{corbett1994knowledge,lee2012impact} that may muddle our understanding of equity for ND students; 3) several technical hurdles need to be overcome to incorporate our approach into BKT; and 4) although Deep Knowledge Tracing (DKT) \cite{piech2015deep} models can probably learn latent representations that correlate to DRT preferences, this is at the expense of control and interpretation of the effects.

\subsection{Simulations}
\label{sec:methods:sim}

In the simulated dataset, we assume that the ability of ND and NT students are drawn from the same distribution, meaning that ability and NDCs are independent. The NDCs considered in this initial work are dyslexia, dyscalculia, and SPD. These chosen conditions reflect a wide range of effects from different delivery and response types, but this work could be applied to others. 

\begin{table}[t]
    \small
    \centering
    \caption{Description of parameter distributions used to generate synthetic dataset. Each parameter was randomly assigned from distributions. Users are given an intrinsic ability and the possibility of one or more ND conditions. Items are assigned a difficulty, discrimination, guessing, subject, content type, information density, delivery type and response type. Information density describes how much information is provided---0.1 represents only a few words, 1 is a large block of text---designed to reflect how clearly an item is presented.}
    \begin{tabular}{M{2.5cm}M{2.5cm}M{2.3cm}}
        \toprule
        Parameter           & Value (Range) & Probability\\
        \midrule
        Ability             & $(-\infty, \infty)$    & $\mathcal{N}(0, 1)$ \\
        
        ND condition        & Dyslexia, Dyscalculia, SPD & 0.1, 0.06, 0.11 \\
        Difficulty          & (-2, 2)       & uniform \\
        Discrimination      & (0.5, 4)      & uniform \\
        Guessing            & (0, 0.15)     & uniform \\
        Subject             & Maths, English & 0.5, 0.5 \\
        Content type        & Letter, Digit, Both & M: 0.1, 0.5, 0.6, E: 1, 0, 0 \\
        No. attempts        & 20            & fixed \\
        Info. density       & $(0.1, 1)$     & $\mathcal{N}(0.35, 0.15)$ \\
        Delivery type       & Read, Listen, Both &  0.3, 0.3, 0.4\\
        Response type       & Written, Speak, Click Picture, Click Read & 0.4, 0.2, 0.2, 0.2 \\
        \bottomrule 
    \end{tabular}

    \label{tab:params_syn_data}
\end{table}

Datasets are created based on the parameters outlined in \Cref{tab:params_syn_data}. These features contribute to the estimation of LQFs and the probability a user will respond to an item. For example, a dyslexic user's learning quality is impacted by delivery types involving reading letters, and response types involving reading letters to click the correct answer(s) or writing an answer that includes letters. A dyscalculic user is affected by delivery and response types involving digits. And someone with SPD is impacted when the delivery involves both reading and listening with either letters and/or digits, as this can cause sensory overload \cite{papathoma2020guidance}. 

\begin{figure}[b]
    \centering
    \includegraphics[width=0.9\linewidth]{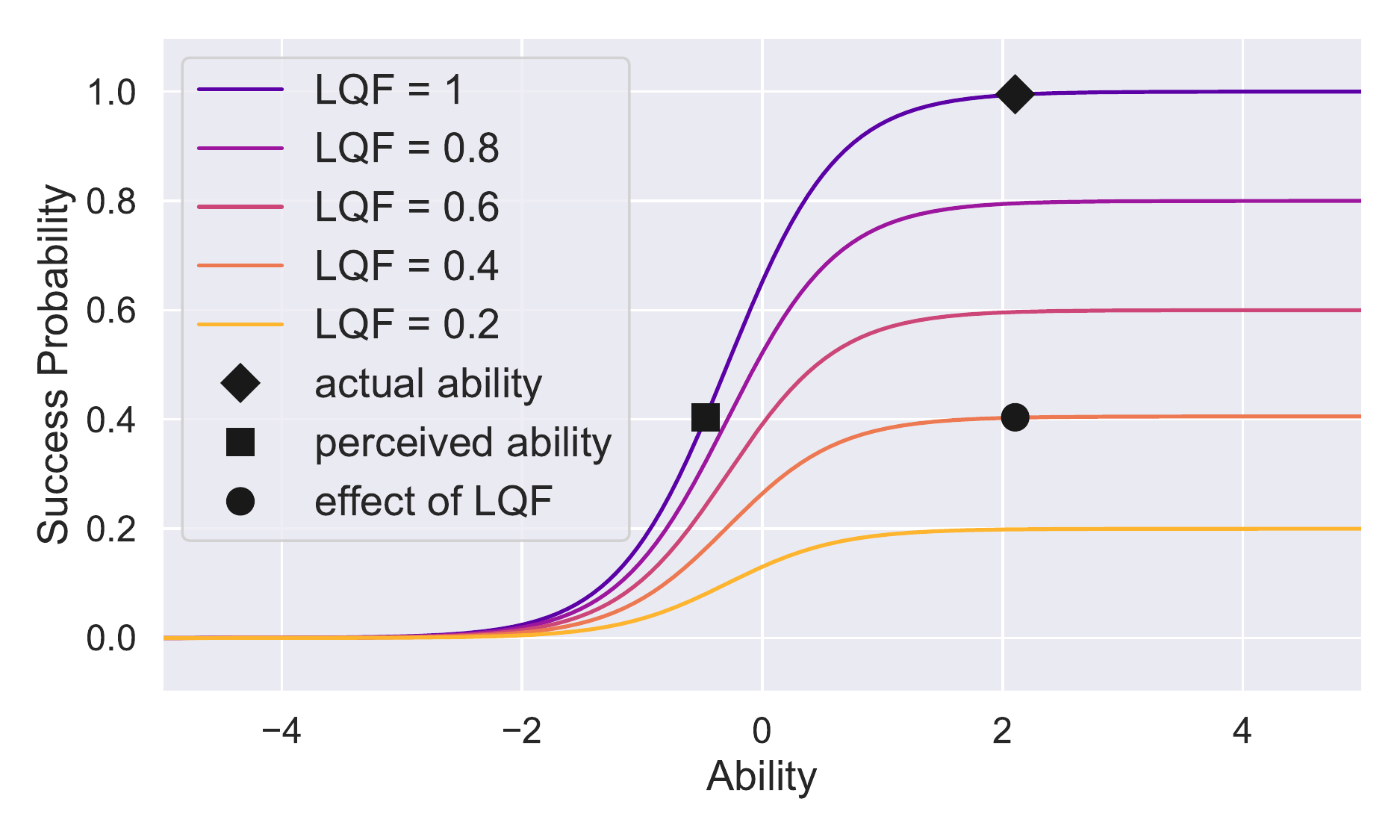}
    \caption{This image shows that LQFs can make the perceived ability of an affected student much lower than their true unobserved ability. $\lozenge$ shows a student’s true ability, $\circ$ shows the impact of low LQFs, and $\square$ shows the perceived ability if LQF is not considered.}
    \Description[Effect of various LQFs on success probabilities. ]{This image shows that LQFs can make the perceived ability of an affected student much lower than their true unobserved ability. $\lozenge$ shows a student’s true ability, $\circ$ shows the impact of low LQFs, and $\square$ shows the perceived ability if LQF is not considered.}
    \label{fig:impedance_item_curve}
\end{figure}

Collectively, these features are used to describe the suitability of DRT to a variety of NDCs, which we now relate back to \Cref{eq:zim}. If a poorly chosen DRT is selected for a ND student, this will result in poor learning opportunities due to a low LQF (\ie large $\pi$). However, if a suitable DRT is selected for a student, the suitability is reflected in higher LQFs.  
Synthesising datasets that adapt to DRTs and NDCs requires specification of the weight vectors to adapt $\pi$ to context (\eg `reading' should increase $\pi$ / reduce LQF for dyslexic but not for dyscalculic students). Although specification of weight vectors is a subjective process, it allows us to express our intuition and instincts about the influential pathways. These are fully described in our implementation\footnote{\url{ github.com/niall-twomey/zero-inflated-learner-models }}.

The effect of LQFs on an item's characteristic curve can be seen in \Cref{fig:impedance_item_curve}. As the LQF decreases, the upper asympotote is reduced, indicating that their opportunity to learn from the interaction is compromised. With this, we interpret LQF as a measure of the contextual inequity.

\section{Results and Discussion}
\label{sec:results}

There are four main questions we want to explore in this work: 1) how much are ND users learning opportunities impacted by poor DRTs; 2) is it possible to identify users with potential NDC based on their performance on items with a range of DRTs; 3) is it possible to estimate user true abilities, accounting for any poor performance due to other factors; and 4) can student learning quality and success be improved through active selection of DRTs?

\begin{figure}[t]
    \centering
    \includegraphics[width=0.9\linewidth]{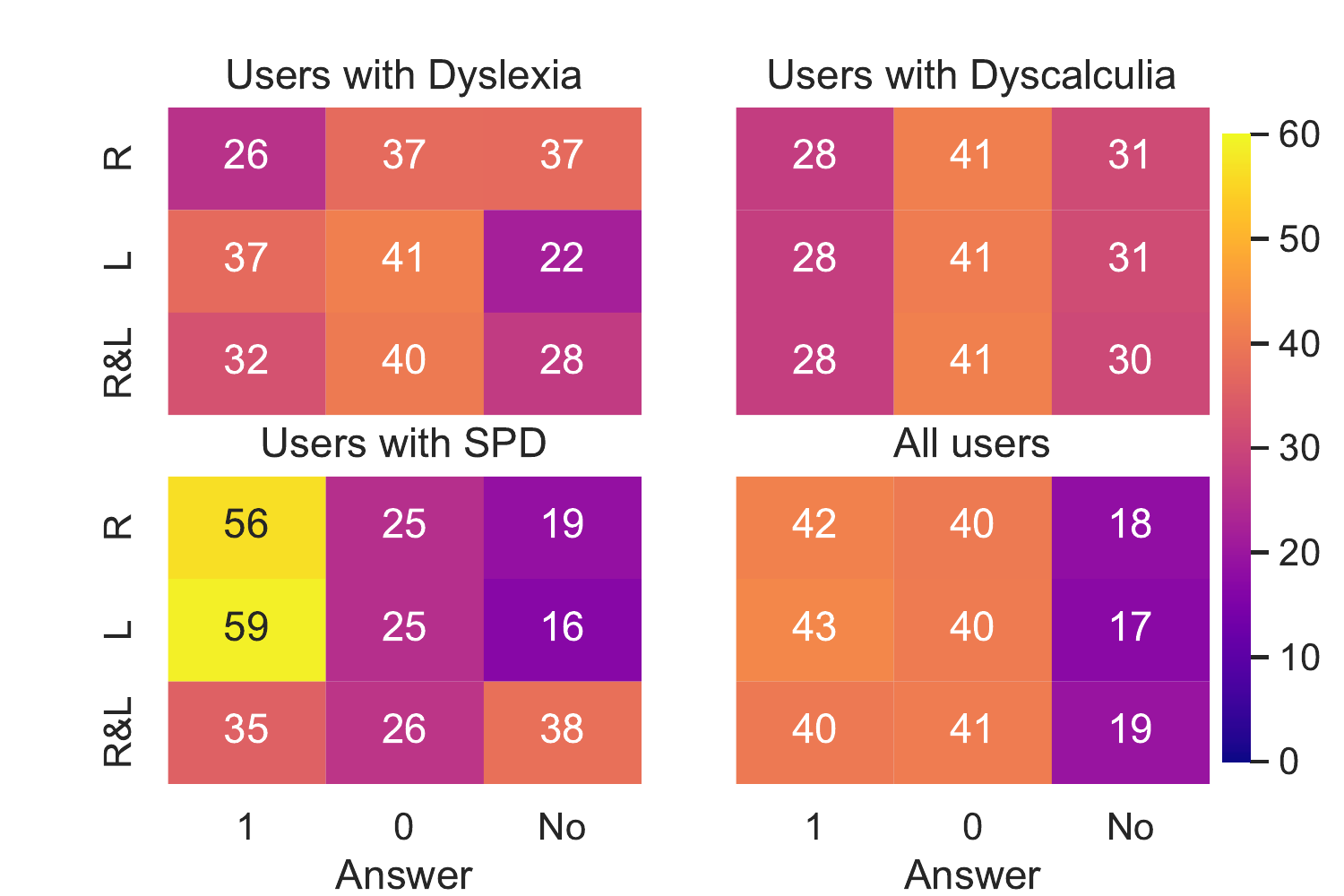}
    \caption{Comparison of attempt outcomes for each NDC (dyslexia: upper left; dyscalculia: upper right; SPD: lower left; all: lower right) when a single delivery type is used for all items (R and L correspond to `read' and `listen' respectively).}
    \Description[Comparison of attempt outcomes for each NDC]{Comparison of attempt outcomes for each NDC (dyslexia: upper left; dyscalculia: upper right; SPD: lower left; all: lower right) when a single delivery type is used for all items (R and L correspond to `read' and `listen' respectively).}
    \label{fig:delivery_heatmaps}
\end{figure}

\subsection{How are ND users impacted?}
\Cref{fig:delivery_heatmaps} shows how ND student performance is affected if a learning environment only delivers information in a single format. 
Across the full neurodiverse population, the mean performance is approximately the same for all learning material formats. There are also no observable differences in performance for users with dyscalculia. However, for users with dyslexia or SPD there are noticeable differences. For users with dyslexia, they answer 6--11\% more attempts correctly and are able to attempt 9--15\% more items when the item has a listening component. For users with SPD, they answer an item correctly, and are able to attempt, 19--24\% more attempts when the item is only delivered in one format compared to multiple formats. The probability of a user succeeding at an item is can be drastically effected by a poor learning quality.

\begin{figure*}
    \centering
    \includegraphics[width=\textwidth]{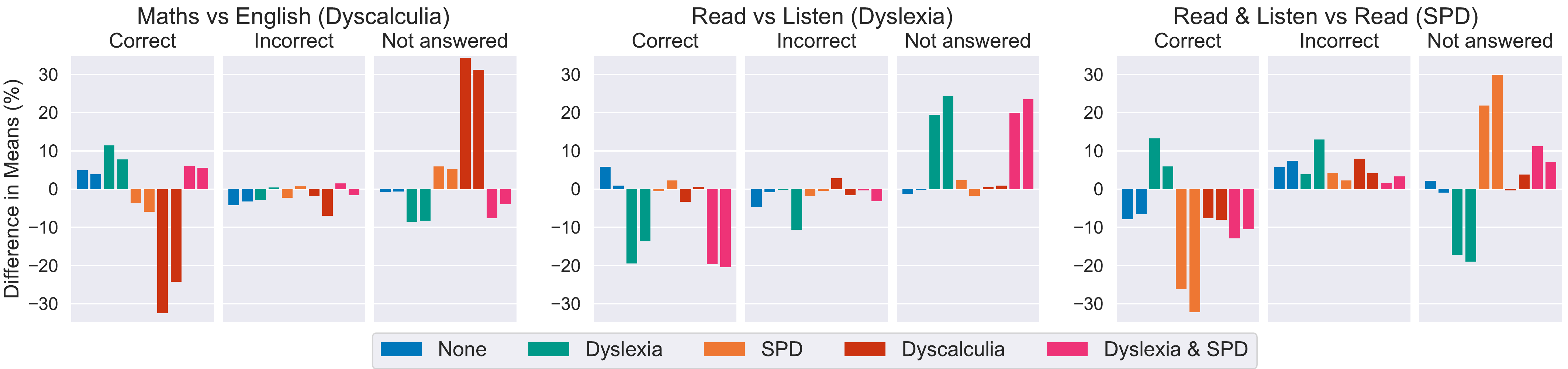}
    \caption{Performance differences for selected students across subject- and NDC-oriented contexts. Bar chart colour indicates NDCs. Large positive and negative values in the bar charts indicates that a group has been affected by the context. While every NDC is affected (indicated in parentheses in subfigure captions), no significant effects present for NT students.}
    \Description[Performance differences for selected students]{Performance differences for selected students across subject- and NDC-oriented contexts. Bar chart colour indicates NDCs. Large positive and negative values in the bar charts indicates that a group has been affected by the context. While every NDC is affected (indicated in parentheses in subfigure captions), no significant effects present for NT students.}
    \label{fig:diff_histograms}
\end{figure*}

\subsection{Can NDCs be identified from interactions?}
To investigate if users with a potential NDCs can be identified from the interactions, we have compared individuals mean performance in different subjects and on items with different delivery types (\Cref{fig:diff_histograms}). When Maths and English are compared (\Cref{fig:diff_histograms} left), dyscalculic users have attempted more English items than Maths (large spike on `Not answered'). Additionally, when Maths is attempted, there is a lower success rate than in English (dip in `Correct'). Their performance in terms of `Incorrect' counts in English and Maths are equivalent. However, this tally is achieved with 30\% fewer attempts, indicating poor performance in Maths, further illustrating the effect of their NDC (\ie 10/20 \vs 5/15). The most noticeable effects between read \vs listen DRT (\Cref{fig:diff_histograms} middle) are seen by a clear increase in number of not answered items and decrease in the number of correct answers for `dyslexia' and `dyslexia \& SPD' students. SPD students are unaffected by these DRTs. Comparing the `read \& listen' and `read' delivery types (\Cref{fig:diff_histograms} right), there are features seen with the dyslexia users, as above, but the SPD users now show a significant difference in performance, with large increases on `not answered' and decreases on `correct'. So, by comparing individual students' performance in different subjects and DRTs, it's possible to identify the ND students and their condition. In practices, these comparisons could be used to identify what contexts a student may be struggling with, and additional support they may need.

\subsection{Can a user's true ability be estimated?}

\begin{figure}[t]
    \centering
    \subfloat[IRT ability\label{fig:irt-ability}]{\includegraphics[width=0.49\linewidth]{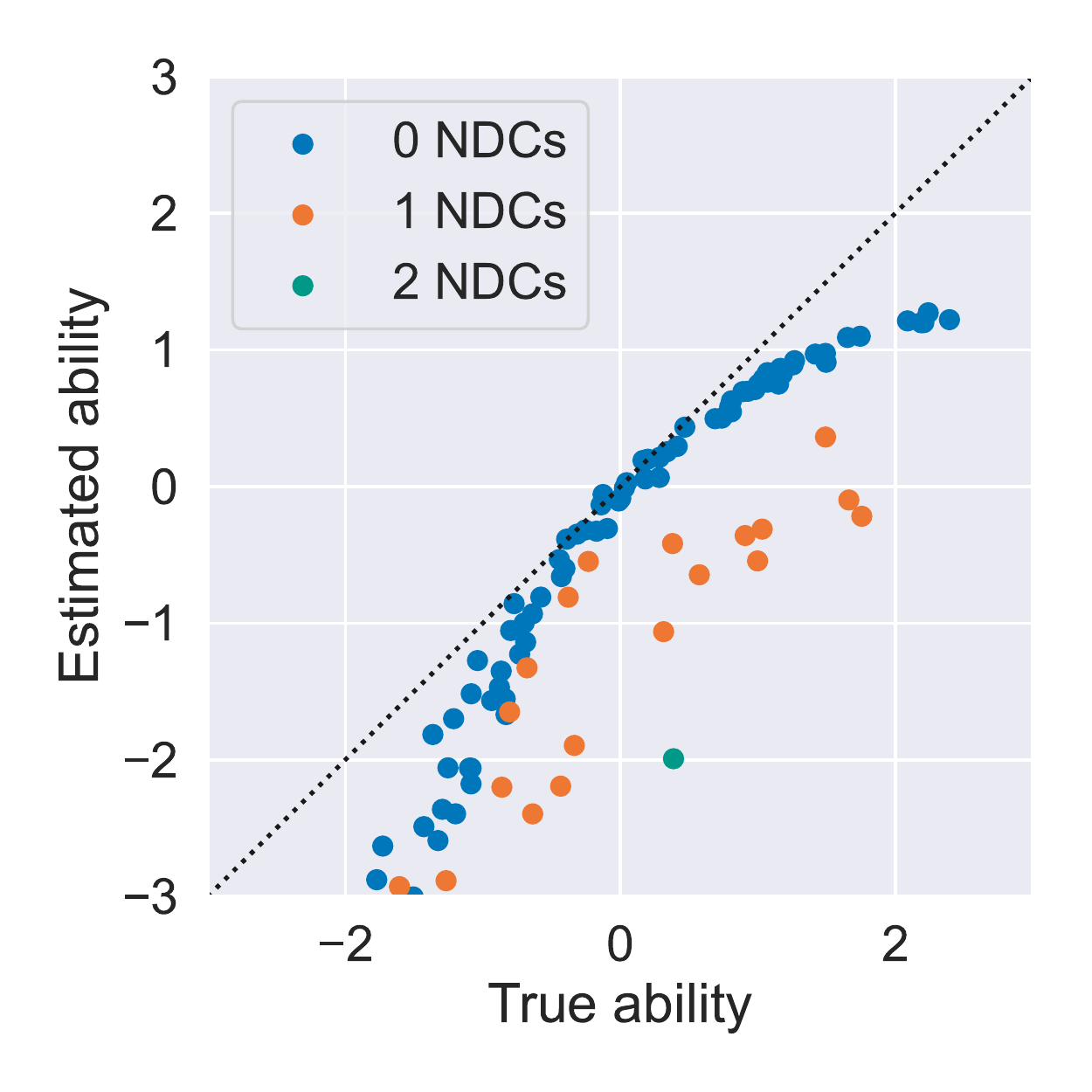}} \hfill
    \subfloat[IRT-ZILM ability\label{fig:zicirt-ability}]{\includegraphics[width=0.49\linewidth]{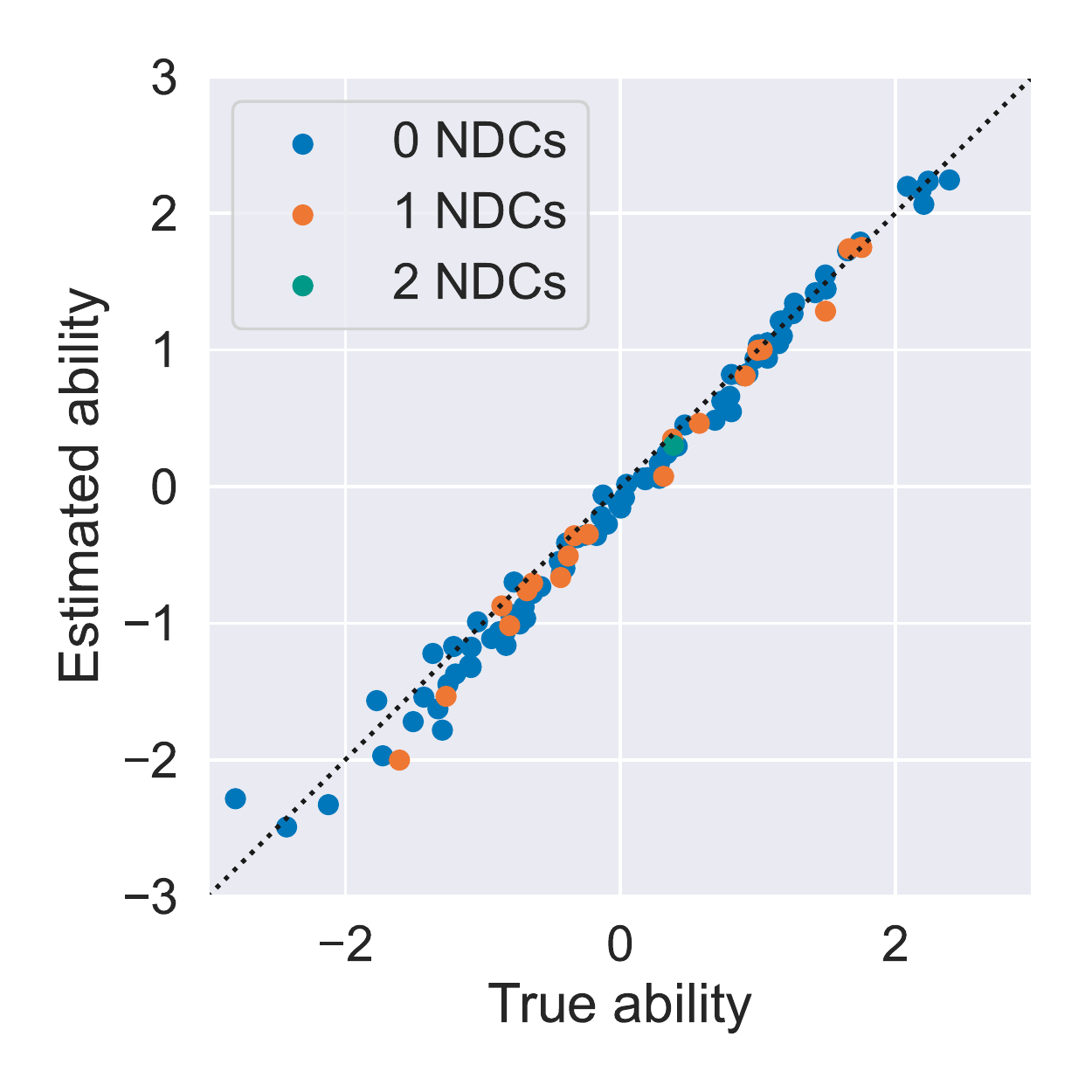}} \hfill
    \Description[Scatter plots of true \vs estimated ability parameters]{Scatter plots of true \vs estimated ability parameters from IRT and IRT-ZILM. Perfect estimation will place all points on diagonal. IRT is biased against ND students, while IRT-ZILM parameter estimation is very reliable.}
    \caption{Scatter plots of true \vs estimated ability parameters from IRT and IRT-ZILM. Perfect estimation will place all points on diagonal. IRT is biased against ND students, while IRT-ZILM parameter estimation is very reliable.}
    \label{fig:ability-recovery}
\end{figure}

One aspect of ensuring each user gets suitable learning material is understanding their true ability. \Cref{fig:ability-recovery} compares the performance of classical IRT and our IRT-ZILM model for parameter recovery. With IRT, most of the ability values are under-estimated, particularly for students with 1 or 2 NDCs (\Cref{fig:irt-ability}). Under-estimated ability makes sense given our expected inflated zero counts. However, the bias of under-estimated ability for ND students is concerning given that ND and NT abilities were drawn from the same distribution. On the other hand, IRT-ZILM is a much better estimator of true abilities (\Cref{fig:zicirt-ability}). Additionally, there is no obvious gap in ability estimates for students with NDCs compared to NT students. \Cref{tab:performance-metrics} summarises the predictive accuracy of the considered models. Although the performance of all models is approximately equivalent (only small gains for our approach) the lack of distorted recovered parameters may indicate stronger reliability of IRT-ZILM. 

\Cref{tab:correlation} summarise parameter estimation using Pearson and Spearman correlation coefficients, and have included linear KTM \cite{vie2019knowledge} (using contextual features) as another baseline. KTM, like IRT also under-estimates ND ability, and IRT-ZILM is a significantly better estimator of the true parameters.

\begin{table}[]
\small
\centering
\caption{Predictive test metrics. Similar performance obtained with all models, though IRT-ZILM is slightly more performant than baselines.}
\label{tab:performance-metrics}
\begin{tabular}{cccc}
\toprule
Metric      & IRT   & KTM & IRT-ZILM \\
\midrule
Accuracy    & 0.734 & 0.742 & \textbf{0.753}   \\
F1          & 0.559 & 0.567 & \textbf{0.583}   \\
NLL         & 0.513 & 0.499 & \textbf{0.494}   \\
Brier Score & 0.170 & 0.166 & \textbf{0.163}   \\
\bottomrule
\end{tabular}
\end{table}

\begin{table}[]
\small
\centering
\caption{Pearson and Spearman correlation coefficients between true and recovered parameters. Values of 1 indicate perfect matches. IRT-ZILM parameter estimation is the most accurate across both metrics for all parameters. }
\label{tab:correlation}
\begin{tabular}{ccccccc}
\toprule
               & \multicolumn{3}{c}{Pearson} & \multicolumn{3}{c}{Spearman} \\
               \cmidrule(lr){2-4} \cmidrule(lr){5-7}
      & IRT & KTM & IRT-ZILM & IRT & KTM & IRT-ZILM  \\
\midrule
Ab      & 0.839 & 0.955 & \textbf{0.993} & 0.929 & 0.966 & \textbf{0.996}  \\
Diff         & 0.394 & 0.686 & \textbf{0.953} & 0.413 & 0.707 & \textbf{0.954}  \\
Disc         & 0.270 & 0.544 & \textbf{0.932} & 0.234 & 0.610 & \textbf{0.942}  \\
\bottomrule
\end{tabular}
\end{table}

\subsection{Can learning quality be improved?}
We explore the effect of actively selecting DRTs to improve LQFs and the number of successful learning attempts for ND students in \Cref{tab:attempt-gap}. The table shows the potential that selecting the most suitable DRT can have on learning quality, with large lifts on students with 1 or 2 NDCs.

\begin{table}[h]
\small
\centering
\caption{Increase (and decrease) of learning opportunities obtained with active (and adversarial) DRT selection.}
\label{tab:attempt-gap}
\begin{tabular}{ccc}
\toprule
                   & 1 NDC & 2 NDCs \\
\midrule
Baseline           & 0.391 & 0.123 \\
\midrule
Lift               & 1.432 $\uparrow$   & 1.898 $\uparrow$ \\
Drop               & 0.248 $\downarrow$ & 0.014 $\downarrow$ \\
\bottomrule
\end{tabular}
\end{table}

\subsection{How can this model be applied?}

As already discussed, comparing user interactions in different contexts can identify students who may need additional support in specific areas. Often, high achieving ND students needs can be overlooked since their performance doesn't tend to require interventions. 
With IRT-ZILM, support/adaptions can be put in place early to enable them to reach their full potential since this model is less susceptable to the biases of traditional LMs. IRT-ZILM can be used to better estimate a students true ability, by adapting it to contexts and underestimating their DRTs preferences. This can help identify and explain causes for underperforming students. By understanding which DRTs a student struggles to engage with, alternative items can be provided to help them reach their full potential. These insights can also be used by teachers to explore if the DRTs of their content can be expanded to create an accessible learning environment for all. Education traditionally has taken a one size fits all approach. By harnessing models that incorporate contextual understanding, learning can be tailored to each student, reaching many of those who may previously have felt dejected in learning, as their needs weren't being met.

\section{Conclusions}
\label{sec:conclusions}

Our application of zero-inflated models in learning contexts offers a rich simulation environment of neurodivergent conditions in question answering settings, unbiased evaluations of neurodivergent learners, encourages increased learning quality, and more reliably recovers unbiased ability parameters. On the basis of our successful results we believe that further study and exploration of zero-inflated learner models can yield an inclusive framework for equitable, explainable, and reliable learner models in diverse educational data mining contexts. Future work will expand on the experimentation to new contexts, and the model to new domains.

\balance{}
\clearpage     
\bibliographystyle{abbrv}
{
\scriptsize
\bibliography{main}
}

\newpage{}

\section*{Appendix}
\label{sec:appendix}

This section gives supplementary details of our proposed model, see \cref{sec:methods:irtzilm}. 

\subsection*{Delivery and Response Weakening} 
\label{sec:methods:wsml}
We adapt learner models for NDCs by taking inspiration from techniques used in Weakly Supervised Machine Learning(WSML) \cite{zhou2018brief}. Our approach is to model the interplay between item DRTs and NDCs. Let a binary random variable be drawn from a Bernoulli distribution, $y \sim \text{Ber}(p)$, and let us assume that a label flipping process acts upon $y$ and this results in observations of the corrupted labels, $\tilde{y}$. The mixing matrix, $M$, is defined as follows: 
$$
\begin{pmatrix}
1 - q_0& q_1\\
q_0& 1 - q_1
\end{pmatrix}=\begin{pmatrix}
Pr(\tilde{Y}=1  \vert  Y=1)& Pr(\tilde{Y}=1  \vert  Y=0)\\ 
Pr(\tilde{Y}=0  \vert  Y=1)& Pr(\tilde{Y}=0  \vert  Y=0)
\end{pmatrix}
$$
The $q_{\tilde{y}}$ variables can be selected using prior knowledge and assumptions on the data distributions \cite{menon2015learning,perello2020recycling}. In our setting, we are interested particularly in the contexts when learning of ND students is being sabotaged by the environment, \ie $q_0$. We therefore model $q_0$ (previously introduced as a global parameter) and parameterise it with ND, LQF and interaction features.

\subsection*{IRT-based Zero-Inflated Learner Model}

Our IRT-ZILM merges LMs and ZILM as follows: 
\begin{align}
    \label{eq:irtzilm}
    Pr(Y=y\mid \mathbf{x})=\begin{cases}
    \pi(\mathbf{x}_\pi) + (1 - \pi(\mathbf{x}_\pi)) (1 - p(\mathbf{x}_{p}))& \text{if } y = 0\\
    (1 - \pi(\mathbf{x}_\pi)) p(\mathbf{x}_p)              & \text{if } y=1
\end{cases}\nonumber
\end{align}
where $\pi$ and $p$ from \cref{eq:zim} are now functions leveraging ND/LQF/content features ($\mathbf{x}_\pi$) and LM/collaborative features ($\mathbf{x}_p$).

By separating the functional contribution of confounders ($\pi$) and ability ($p$) in IRT-ZILM, we hope to unambiguously decouple these aspects from each other and improve interpretability and explainability. The model is learnt by gradient descent of negative log likelihood of the training data to optimise all parameters. In WSML it is common to learn in a two-step process, for example, by iteratively fixing and optimising IRT and weak label weights \cite{perello2020recycling}. 

ZIMs have been used to account for excess zeros in many counting tasks using Poisson and negative binomial models \cite{lambert1992zero,wang2010irt,smits2020study,magnus2018zero}, and in learning analytics as statistical counting models in self-regulated learning \cite{greene2011analysis}.  An important property of statistical models is identifiability as it allows for the precise estimation of the values of its parameters \cite[Sec 4.5]{gelman2006data}. Parallel theoretical analysis has considered identifiability of the counting model parameters \cite{roemmele2019flexible} and the mixture components \cite{li2012identifiability}. It is worth noting that IRT also suffers from identifiability problems (\cf \cite[p.6]{chen2021item} and \cite[Sec 14.]{gelman2006data}) but using priors or regularisation can alleviate these. 

As far as we are aware, this is the first work to incorporate ZIM in this manner. Choosing IRT as the base LM in IRT-ZILM over alternative options is motivated for several reasons. Firstly, IRT is well-understood and simple to interpret, and using this model as a platform to demonstrate new properties of equity in this early work carries the same benefits. Secondly, BKT is known to have over- and under-estimation problems \cite{corbett1994knowledge,lee2012impact} which may muddle our understanding of equity for ND students. Additionally, several technical hurdles need to be overcome, notably adaptation for contextualised individualisation in mixed graphs. Finally, although DKT \cite{piech2015deep} models can probably learn latent representations that correlate to DRT preferences, this is at the expense of control and interpretation of the effects.

\subsection*{Extra Results}

\begin{figure}[h]
    \centering
    \includegraphics[width=0.9\linewidth]{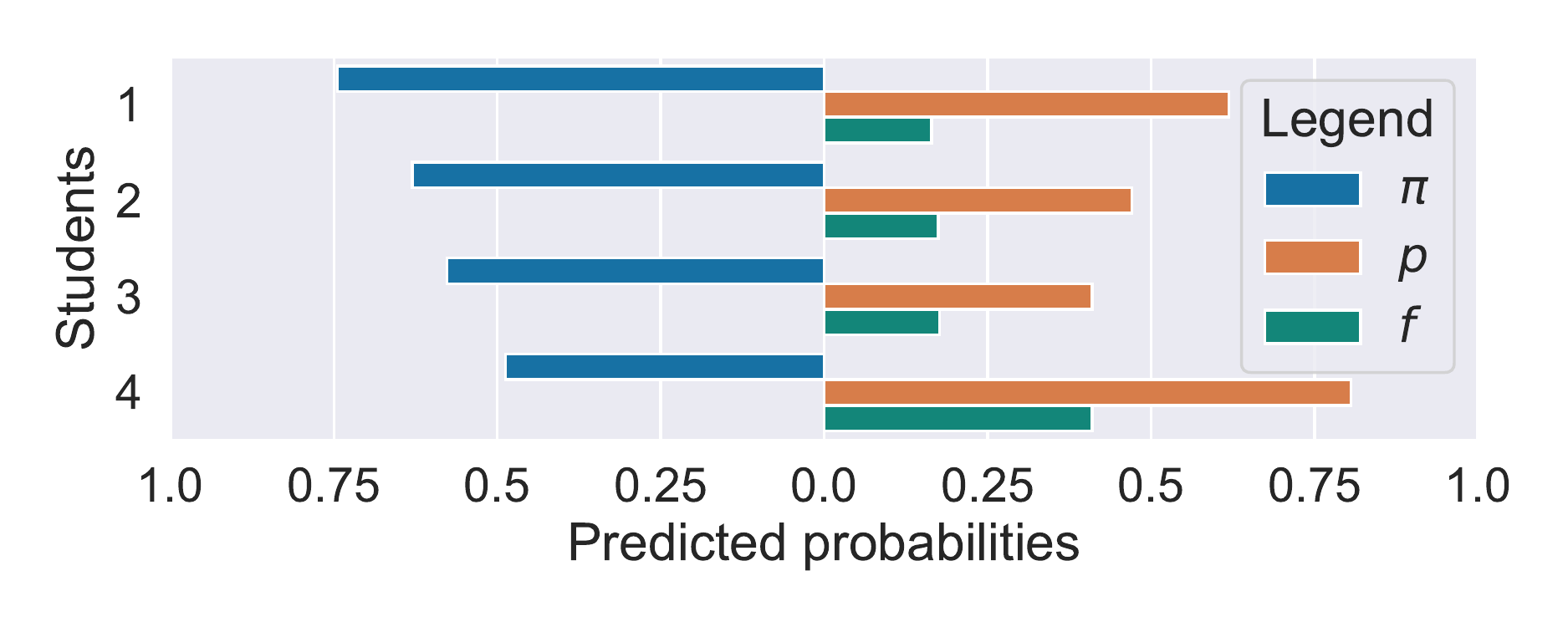}
    \caption{IRT predictions ($f$) combining with complementary LQFs ($\pi$) producing IRT-ZILM predictions ($p$).}
    \Description[Breakdown of IRT and IRT-ZILM predictions.]{IRT predictions ($f$) combining with complementary LQFs ($\pi$) producing IRT-ZILM predictions ($p$).}
    \label{fig:predictive}
\end{figure}

\Cref{fig:predictive} shows this effect for four user/item pairs. For example, the first student should be $60\%$ (orange) successful on this item, however, their LQF is 0.25 (blue), so their success rate drops to $15\%$ (green). Therefore, LQF can be interpreted as a measure of the contextual inequity in these settings.

Although the purpose of this research is to provide equitable estimates of student ability and to provide enabling technology that selects the most appropriate DRT for students, we note that we may also identify students that need additional support in specific areas by recognising potentially unidentified NDCs. We can approach this by creating two models: let $\mathcal{M}_0$ be the model for a student's reported NDC state (the `null' model), and let $\mathcal{M}_1$ be a model trained on data assuming an alternative NDC state (the `alternative' model). Since we have already shown that metrics and likelihood is improved with IRT-ZILM, a statistical hypothesis test can be performed on both likelihoods to determine whether the null or alternative NDC offers a better explanation of data. We leave further elaboration of this approach as future work since it is outside the scope of our direct objectives.

\begin{figure}[h]
    \centering
    \subfloat[IRT difficulty\label{fig:irt-difficulty}]{\includegraphics[width=0.5\linewidth]{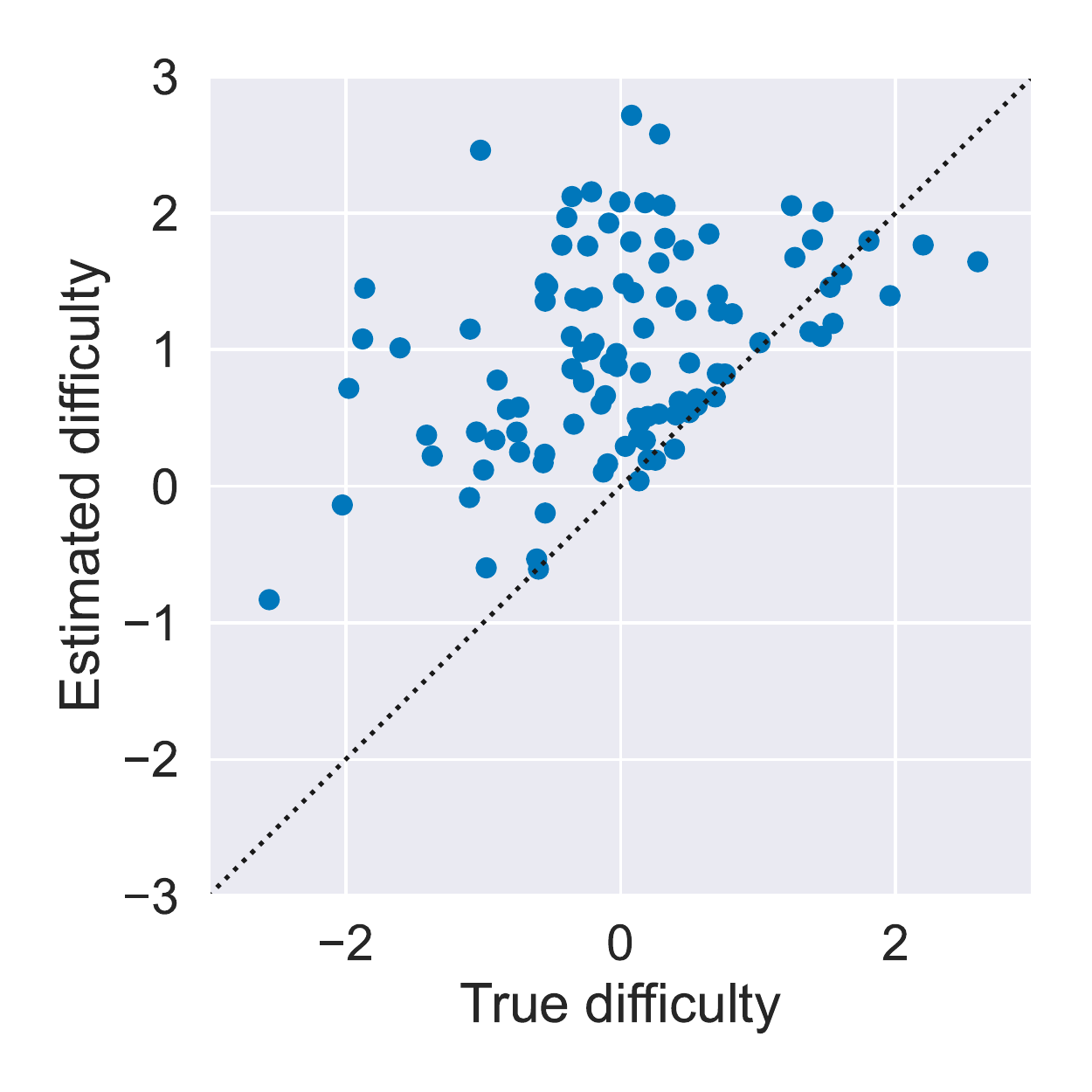}} \hfill
    \subfloat[IRT-ZILM difficulty\label{fig:zicirt-difficulty}]{\includegraphics[width=0.5\linewidth]{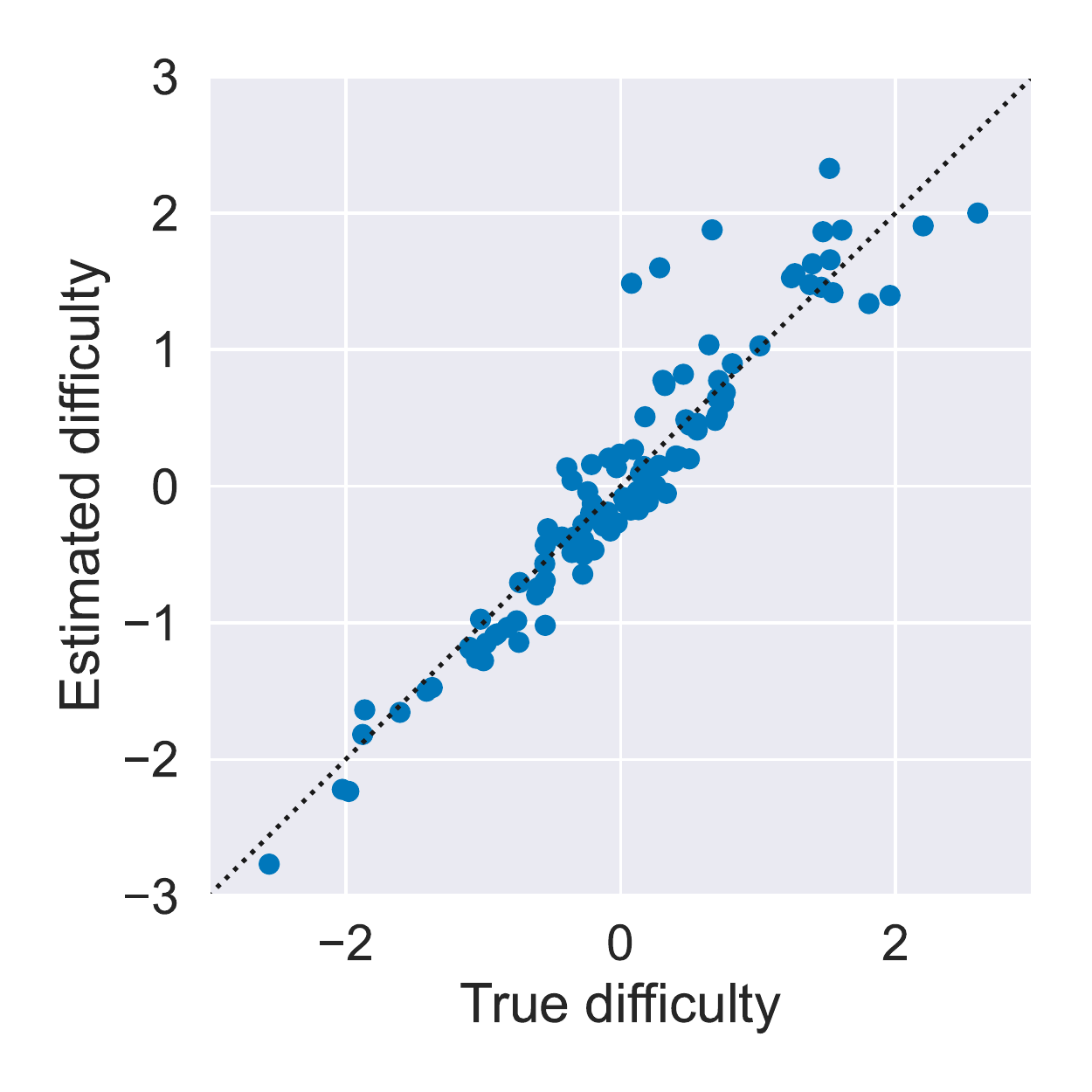}}
    \caption{Scatter plots of true \vs estimated difficulty parameters from IRT and IRT-ZILM. Perfect estimation will place all points on diagonal. Estimation from IRT-ZILM is significantly more accurate than IRT.}
    \Description[Scatter plots of true \vs estimated difficulty parameters]{Scatter plots of true \vs estimated difficulty parameters from IRT and IRT-ZILM. Perfect estimation will place all points on diagonal. Estimation from IRT-ZILM is significantly more accurate than IRT.}
    \label{fig:difficulty-recovery}
\end{figure}

\end{document}